\documentclass[twocolumn,superscriptaddress,secnumarabic,amssymb, nobibnotes, aps, bm,pre,floatfix]{revtex4-1}

\usepackage{lineno}
\usepackage{amsmath,bm,graphicx}
\graphicspath{{Figs/}}
\usepackage[usenames,dvipsnames]{color}
\definecolor{oceanboatblue}{rgb}{0.0, 0.47, 0.75}
\definecolor{tinekegreen}{RGB}{22,140,22}
\definecolor{orange}{rgb}{1,0.5,0}
\definecolor{goodgreen}{rgb}{0.1,0.5,0}
\definecolor{goodred}{rgb}{0.7,0,0}
\usepackage[colorlinks,urlcolor=goodgreen,citecolor=blue,linkcolor=goodred]{hyperref}

\newcommand{\dario}[1]{\textcolor{oceanboatblue}{#1}}

\newcommand{\reyes}[1]{\textcolor{orange}{#1}}

\usepackage[normalem]{ulem}

\newcommand{\comment}[1]{}
\begin{document}

%\title{Co-existence of topological and trivial states in Graphene with Kane-Mele spin-orbit interaction}
\title{Volkov-Pankratov states in topological graphene nanoribbons}
\author{Tineke L. van den Berg}%
\altaffiliation{These authors contributed equally}
\email{tineke.vandenberg@dipc.org}
\affiliation{Donostia International Physics Center (DIPC), 20018 Donostia--San Sebasti\'an, Spain}

\author{Alessandro De Martino} 
\altaffiliation{These authors contributed equally}
\email{ademarti@city.ac.uk}
\affiliation{Department of Mathematics, City, University of London, London EC1V 0HB, United Kingdom}

\author{M. Reyes Calvo}%
%\affiliation{CIC nanoGUNE, 20018 Donostia -- San Sebasti\'an, Spain}
%\affiliation{IKERBASQUE, Basque Foundation for Science, 48013 Bilbao, Basque Country,  Spain}
\affiliation{Departamento de Fisica Aplicada, Universidad de Alicante, 03690 Alicante, Spain}

\author{Dario Bercioux}
\email{dario.bercioux@dipc.org}
\affiliation{Donostia International Physics Center (DIPC), 20018 Donostia--San Sebasti\'an, Spain}
\affiliation{IKERBASQUE, Basque Foundation of Science, 48011 Bilbao, Spain}

\begin{abstract} % should here be ~5 % of the article ; < 500 words

In topological systems, a modulation in the gap onset near interfaces can lead to the appearance of massive edge states, 
as were first described by Volkov and Pankratov. In this work, we study graphene nanoribbons in the presence of 
intrinsic spin-orbit coupling smoothly modulated near the system edges. We show  that this space modulation leads to the appearance of 
Volkov-Pankratov states, in addition to the topologically protected ones. We obtain this result by means of two complementary methods,  
one based on the effective low-energy Dirac equation description and the other on a fully numerical tight-binding approach, finding excellent agreement 
between the two.  We then show how transport measurements might reveal the presence of Volkov-Pankratov states, 
and discuss possible graphene-like structures in which such states might be observed.
\end{abstract}
\maketitle

%\tableofcontents
%\linenumbers

%%%%%%%%%%%%%%%%%%%%%%
\section{Introduction}
%%%%%%%%%%%%%%%%%%%%%%

Graphene was the first material theoretically predicted to be a Quantum Spin Hall (QSH) insulator.  In the proposal by Kane and Mele~\cite{Kane_2005A,Kane_2005B}, the intrinsic spin-orbit coupling (SOC) opens a topological gap in the energy dispersion of the bulk system, and edge states appear in nanoribbons due to the bulk-boundary correspondence. While possible signatures of a topological gap in graphene have been recently reported~\cite{Sichau_2019}, the minute size of the spin-orbit gap in pristine graphene ($\approx 25~\mu$eV) complicates the observation and application of the promising electronic and spin properties of the QSH edge states. Two different approaches have mainly been followed to overcome this limitation: a) Find ways to induce a stronger SOC in graphene, for example by depositing heavy adatoms on the graphene surface~\cite{Weeks_2011,Swartz_2013,Jia_2015,Chandni_2015,Hatsuda_2018} or by proximity to materials with much stronger SOC than carbon such as transition metal dichalcogenides (TMDs) \cite{Gmitra_2015,Wang_2015,Wang_2016,Frank_2018,Wakamura_2019,island2019spin,Tiwari_2020}. b) To grow graphene-like honeycomb structures made of heavier elements in groups IV and V \cite{Liu_2011,Molle_2017,reis2017bismuthene,deng2018_stanene_sts,Yang_2012,Sabater_2013,Drozdov_2014,Li_2018,Pulkin_2019}.
While the experimental realization of QSH physics in these systems seems challenging, and even though there is so far limited evidence for the existence of protected edge states \cite{island2019spin}, the advances in the artificially induced SOC in graphene are promising~\cite{Wakamura_2018,Wakamura_2019}. 

The experimental approaches described above are likely to result into an inhomogeneous distribution of the strength of the induced SOC, 
with larger inhomogeneity near the sample edges. We therefore investigate, both analytically and numerically, the effects of a
reduction of the SOC near the edges in wide graphene nanoribbons, both of zigzag and armchair type. Intrinsic SOC leads to 
band inversion at the $\mathbf{K}$ and $\mathbf{K'}$ points, and due to the smooth SOC modulation at the edges, new massive edge states appear in addition to the topologically protected ones. Such massive edge states were first described by Volkov and Pankratov~\cite{Volkov_1985,Pankratov_1987,Pankratov_2018}, and are therefore refered to as Volkov-Pankratov (VP) states. Recently, VP states have attracted attention again in the context of topological insulators (TIs)~\cite{Inhofer_2017,Tchoumakov_2017,Mahler_2019,Lu_2019,Vandenberg2020} and topological superconductors~\cite{Alspaugh_2020}. Although VP states in TIs are not topologically protected, they are of topological origin, because they result from the band inversion between a topological and a trivial material~\cite{Tchoumakov_2017}.  Three-dimensional TIs with band inversion at the $\Gamma$-point were investigated both theoretically, 
within an effective linear in momentum model~\cite{Tchoumakov_2017,Lu_2019}, as well as experimentally, 
in HgTe/CdTe heterojunctions~\cite{Inhofer_2017,Mahler_2019}. 
Other studies focused on two-dimensional quantum wells within the Bernevig-Hughes-Zhang model~\cite{Vandenberg2020}. 
In these works the VP states appeared due to the smooth modulations in the band structure near the edges. Recently, the coexistence of topological and trivial modes was  also reported in 2D TMD, specifically in the 1T' phase of WSe$_2$~\cite{Pulkin_2019}. 

Within our numerical approach, in addition to the spectral properties, we investigate the transport properties of clean and disordered nanoribbons. In general, the 
conductance of the system increases by $4 e^2/h$ every time a new VP band opens to conduction. In disordered ribbons, the opening of 
a new conduction channel via a VP state is accompanied by  the appearance of a dip in conductance. 
These dips resemble the ones observed in quasi-one dimensional quantum wires in the presence of an attractive impurity~\cite{Bagwell_1990}. 
In this context, the decrease in the conductance is due to the coupling of propagating modes to quasi-bound states in the scattering region. Therefore, the presence of these dips in the conductance indicates that a new subband is opened, thereby demonstrating the existence of such bands within the topological gap. This hints that the presence of disorder is not detrimental to the detection of VP states in transport experiments.

%\dario{Recently, the coexistence of topological and trivial modes was  also reported in 2D TMD, specifically in the 1T' phase of WSe$_2$, where it presents a strong dependence on the material termination~\cite{Pulkin_2019}.}

This article is organized in the following way. In Sec.~\ref{analytics}, we discuss the analytical solution for the 
edge mode spectrum of a semi-infinite graphene flake with a space modulation of the intrinsic SOC close to the boundary, 
within the long-wavelength approximation. 
In Sec.~\ref{numerics}, we solve for the full spectrum of zigzag and armchair nanoribbons with the modulated intrinsic
SOC using the tight-binding model. In Sec.~\ref{transport}, we analyse the transport properties within the full 
tight-binding description, including the effects of disorder. 
Finally, in Sec.~\ref{conclusions}, we present conclusions and outlook. Some technical details are relegated 
to the three appendices at the end of the paper.

%%%%%%%%%%%%%%%%%%%%%%
\section{Analytical low-energy approach}\label{analytics}
%%%%%%%%%%%%%%%%%%%%%%

In this section we investigate the spectrum of edge states of graphene nanoribbons 
by using the Dirac equation description~\cite{Brey_2006,CastroNeto_2009}. As it is well-known, 
this emerges in the long-wavelength approximation (LWA) of the tight-binding Hamiltonian 
around the Dirac points $\mathbf{K}$ and $\mathbf{K}'$. Within this continuum model,
we account for the presence of non-uniform intrinsic SOC. We focus on the part of the spectrum 
corresponding to states localized at the edges. These states decay exponentially away from the edges in the bulk, 
so if the width of the nanoribbon is much larger than their decay length, the two edges can be treated separately. 
We assume that this is the case, and study a semi-infinite system with a single edge of either zigzag or armchair type. 
Our results below, therefore, apply to {\em wide} nanoribbons, and do not account for finite-size effects. 
%
%
%%%%%%%%%%%%
\begin{figure}
    \centering
    \includegraphics[width=0.51\linewidth]{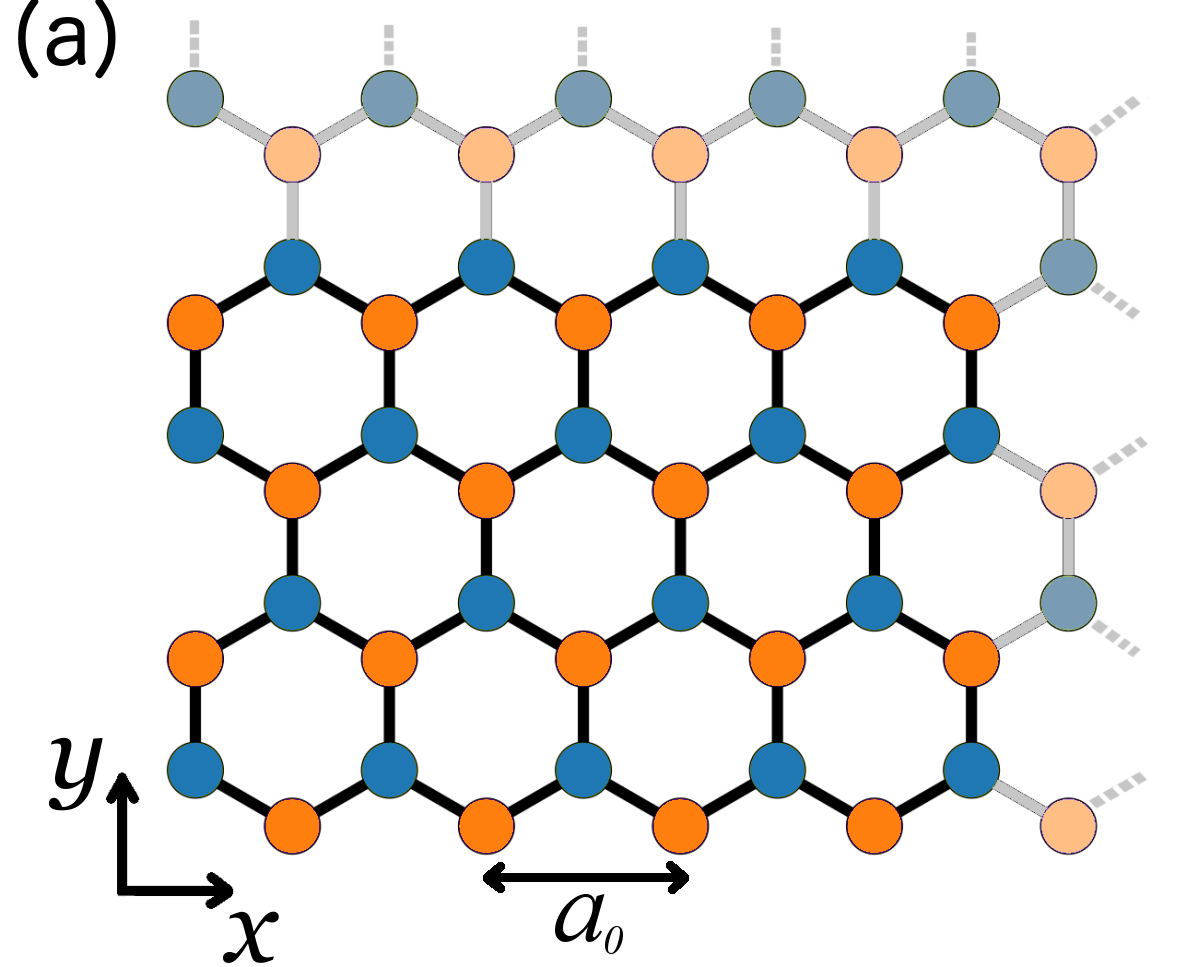}
        \includegraphics[width=0.47\linewidth]{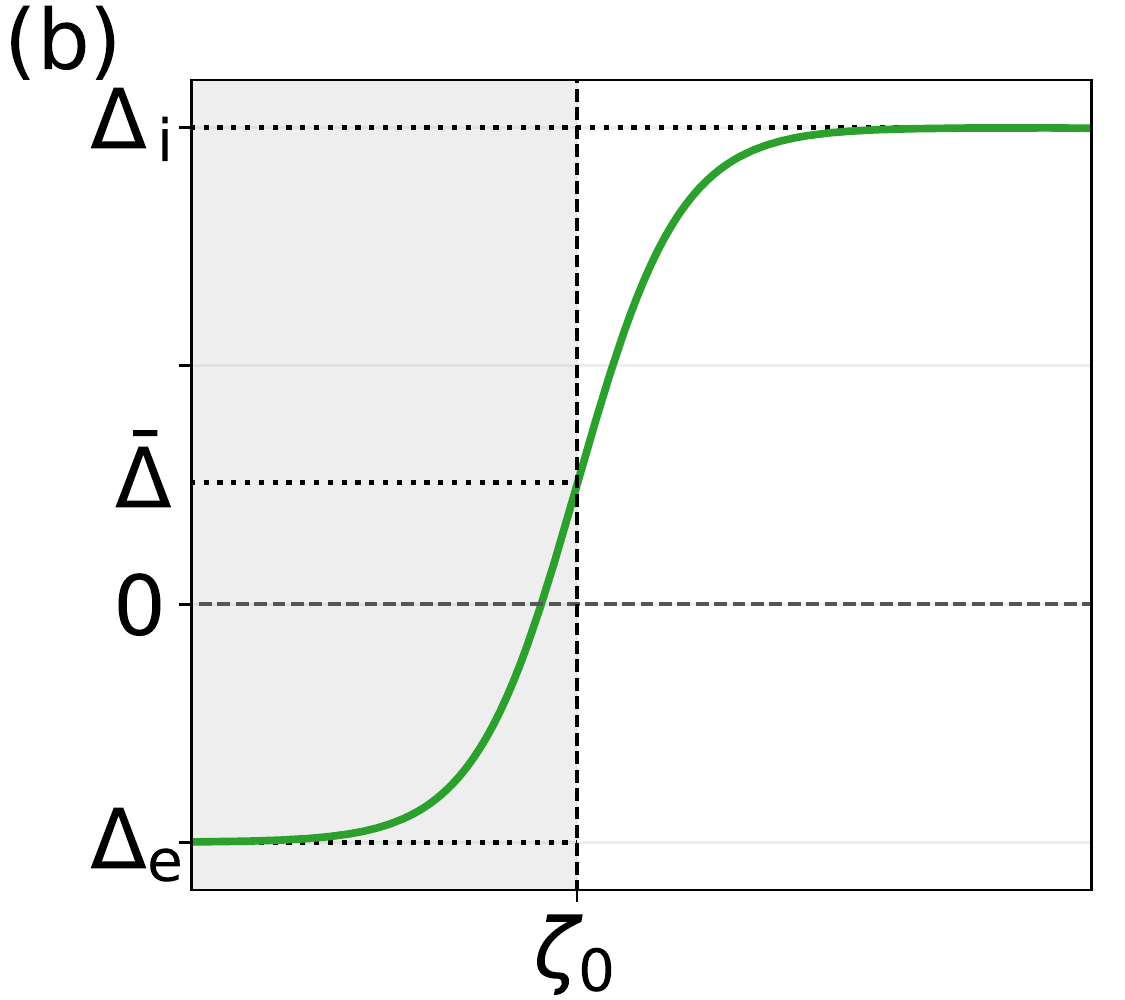}
\caption{\label{fig_system} (a) Sketch of the system with $x$ and $y$ directions. 
In the analytical calculation the system is a semi-infinte half plane. 
In the numerical calculations it has a ribbon geometry, attached to source and drain leads. 
Depending on the orientation, it is a system with either zigzag, or armchair edges. 
(b) Sketch of the profile of the space modulation of the intrinsic SOC 
given by Eq.~\eqref{Delta_modulation}. The system occupies the unshaded region.}
\end{figure}
%%%%%%%%%%%%
%
%
Within the LWA approximation, graphene's effective Hamiltonian is given by
%
%
%
%%%%%%%%%%%%%
\begin{align}\label{LWAHamiltonian}
\mathcal{H}= v_F   (\tau_z \sigma_x \hat p_x + \sigma_y \hat p_y) + 
\Delta  \tau_z \sigma_z  s_z\, ,
\end{align}
%%%%%%%%%%%%
%
%
where $v_F$ denotes graphene's Fermi velocity, $ \bm{p} = -i\hbar \bm{\nabla}$ is the momentum 
operator, and $\{\tau_x,\tau_y,\tau_z\}$/$\{\sigma_x,\sigma_y,\sigma_z\}$/$\{s_x,s_y,s_z\}$
are the Pauli matrices associated to the valley/sublattice/spin degree of freedom, respectively.
Since the Hamiltonian~\eqref{LWAHamiltonian} is diagonal in valley and spin space, 
we will work in a basis of given valley and spin projection: 
%
%
%
%%%%%%%%%%%%%
\begin{align}\label{FullWF}
\Psi_{\tau,s} = \begin{pmatrix}
v_{A ,\tau,s} \\ 
v_{B ,\tau,s}\end{pmatrix}, \quad \tau,s=\pm 1,
\end{align}
%%%%%%%%%%%%%
%
%
%
and we shall omit the valley and spin indices wherever there is no risk of confusion. 
In Eq.~\eqref{LWAHamiltonian}, we include a non-uniform intrinsic SOC with a 
smooth spatial modulation transverse to the boundary:
%
%
%%%%%%%%%%%%%
\begin{align}\label{Delta_modulation}
\Delta (\zeta) &= \bar \Delta  + \Delta_0 \tanh \left( \frac{\zeta-\zeta_0}{\ell } \right).
\end{align}
%%%%%%%%%%%%
%
%
Here, $\zeta$ represents the coordinate in the direction perpendicular to the boundary, 
located at $\zeta=0$, and the system extends on the side of positive $\zeta$.
Equation~\eqref{Delta_modulation} describes a domain-wall profile centered 
at $\zeta=\zeta_0$, with characteristic modulation length $\ell$,
and with asymptotic values given by
%
%
%%%%%%%%%%%%%
\begin{align}
\Delta_\text{i,e} &= \bar \Delta  \pm \Delta_0.
\end{align}
%%%%%%%%%%%%
%
%
Then $\Delta_\text{i}$ represents the SOC deep in the interior of the nanoribbon. 
We assume that the modulation occurs close to the boundary, where the SOC 
reduces to $\bar \Delta$, with $\zeta_0$ of the order of graphene's lattice constant $a_0$,
see Fig.~\ref{fig_system}(b). 

The choice of the hyperbolic tangent profile is convenient because it allows for an exact solution of the 
corresponding Dirac equation~\cite{LandauQM,Tchoumakov_2017}. However, we expect that
the qualitative features of the spectrum do not depend on the detailed shape of the profile, 
as long as the typical length scale $\ell$ of the SOC modulation is large on the scale of 
the lattice constant $a_0$. Since this is also the condition for the validity of the LWA 
employed here, throughout this paper we assume $\ell\gg a_0$.

%%%%
\subsection{Spectral properties of zigzag ribbons}
%%%%
We start by considering a semi-infinite graphene flake extending in the region $y>0$, 
with a zigzag edge along the $x$-axis [see Fig.~\ref{fig_system}(a)],
and SOC profile given by Eq.~\eqref{Delta_modulation} with $\zeta = y$.
By exploiting the translational invariance along the $x$-direction, 
the wave function can be expressed in the form  
%
%
%%%%%%%%%
\begin{align}\label{wfzigzag}
\Psi(x,y)=e^{ik_x x}\psi(y),
\end{align}
%%%%%%%%%
%
%
with $\psi^T(y)=(v_A ,v_B)$.
Then, the Dirac equation reduces to
%
%
%%%%%%%%%
\begin{align}\label{diraczigzag}
\left[ \tau \sigma_x k_x - i \sigma_y \partial_y +  s \tau \Delta(y) \sigma_z\right]\psi=E\psi \,,
\end{align}
%%%%%%%%%
%
%
where we set $\hbar=1$, and measure energies in units of $v_F/\ell$, 
lengths in units of $\ell$, and wave vectors in units of $\ell^{-1}$.
Equation~\eqref{diraczigzag} admits an exact solution \cite{LandauQM,Tchoumakov_2017},
whose derivation is summarized in App.~\ref{analytic_solution}. 
Here, we just present the result. We introduce the notation
%
%
%%%%%%%%%
\begin{align}
\kappa _\text{i/e} = \sqrt{k^2_x+\Delta_\text{i/e}^2-E^2}\,, \qquad 
\bar \kappa  = \frac{\kappa_\text{i}+\kappa_\text{e}}{2}\,.
\end{align}
%%%%%%%%%%%%
%
%
Then, in terms of the new variable $u$ given by
%Using the change of variable $y \rightarrow u$
%
%
%%%%%%%%%
$$
u=\frac{1}{2}\left[1- \tanh (y-y_0) \right],
$$
%%%%%%%%%%%%
%
%
the sublattice amplitudes can be expressed as 
%
%
%%%%%%%%%
\begin{align}
\label{vZigZag}
%v_\text A = u^{\kappa_\text{R}}(1-u)^{\kappa_\text{L}} (w_++w_-)\,,\\
%v_\text B = u^{\kappa_\text{R}}(1-u)^{\kappa_\text{L}} (w_+-w_-)\,, \\
v_\alpha = u^{\kappa_\text{i}/2}(1-u)^{\kappa_\text{e}/2} (w_++ \alpha\, w_-)\,,
\end{align} 
%%%%%%%%%%%%
%
%
with $\alpha=A,B=\pm$. The functions $w_\pm(u)$ 
satisfy a hypergeometric equation (see App.~\ref{analytic_solution}).
Selecting the solution that leads to normalizable states for $y\rightarrow +\infty$
(i.e., $u\rightarrow 0$), we obtain
%
%
%%%%%%%%%
\begin{align}
\label{wZigZag}
%w_+(u) &=A_+\,  F[\bar \kappa +\Delta_0,\bar \kappa -\Delta_0 +1 ; \kappa_\text{R} +1; u],\\
%w_-(u) &=A_-\, F[\bar \kappa-\Delta_0,\bar \kappa +\Delta_0 +1 ; \kappa_\text{R} +1; u], \\
w_\pm (u) &=c_\pm \,  F[\bar \kappa \pm s\tau\Delta_0,\bar \kappa 
\mp  s\tau \Delta_0 +1 ; \kappa_\text{i} +1; u],
\end{align}
%%%%%%%%%%%%
%
%
where $F(a,b;c;z)$ is the  ordinary hypergeometric function \cite{Olver:2010}.
%
%
%%%%%%%%%
Normalizability requires $\kappa_\text{i}>0$, but 
imposes no constraint on $\kappa_\text{e}$. 
Therefore, at fixed $k_x$, the edge modes exist in energy window $|E|<\sqrt{\Delta^2_\text{i}+k_x^2}$.
Since for $y\rightarrow +\infty$ we have $u\sim e^{-2y}$, $\kappa^{-1}_\text{i}$
actually represents the decay length of the corresponding edge mode into the bulk. The relative factor between the two spinor components is 
fixed by the Dirac equation. We find
\begin{align}
\label{coeffZZ}
\frac{c_-}{c_+} = \frac{\kappa_\text{i} + s\tau \Delta_\text{i}}{E+\tau k_x}\, .
%= \frac{\tau k_x-E}{\kappa_\text{i}- s \tau \Delta_\text{i}} 
\end{align}
%%%%%%%%%%%%
%

We now need to impose the appropriate boundary condition. In the case of 
a zigzag edge, the boundary condition requires that one sublattice component of 
the wave function vanishes at the boundary $y=0$, separately for each valley and spin 
\cite{Brey_2006,CastroNeto_2009}:
%
%
%%%%%%%%%
\begin{align}\label{zigzagAb}
v_{\alpha, \tau,s}(0)=0, \quad \tau,s=\pm 1 \, ,
\end{align}
%%%%%%%%%
%
%
where $\alpha =A$ (respectively $\alpha=B$) if the boundary sites belong to 
the $B$ (respectively $A$) sublattice. Using Eqs.~\eqref{vZigZag}, \eqref{wZigZag}, 
and \eqref{coeffZZ}, from Eq.~\eqref{zigzagAb} we obtain
%
%
%%%%%%%%%%%%%%
\begin{align} \label{eq_zigzagFinalEq}
\frac{(E+\tau k_x) F[ \bar \kappa + s\tau\Delta_0, \bar \kappa - s\tau\Delta_0 +1 ; 
\kappa_\text{i} +1;u_0]}
{(\kappa_\text{i}+ s\tau \Delta_\text{i})F[\bar \kappa -s\tau\Delta_0, 
\bar \kappa +s\tau\Delta_0+1 ; \kappa_\text{i} +1; u_0] }\!=\!- \alpha,
\end{align}
%%%%%%%%%%%%
%
%
where  $u_0=\frac{1}{2}(1+\tanh y_0)$. The shift $y_0$ is important for the 
comparison to the numerical tight-binding analysis, in which the SOC profile 
is centered exactly at the edge of the system, i.e., on the first line of 
carbon atoms \cite{Wakabayashi_2010}. In the continuum approach, 
this corresponds to the choice $y_0=a_0/\sqrt{3}$, because $y=0$ corresponds to a 
line of auxiliary sites, where one imposes the vanishing of the wave function.

\begin{figure}
    \centering
    \includegraphics[width=\linewidth]{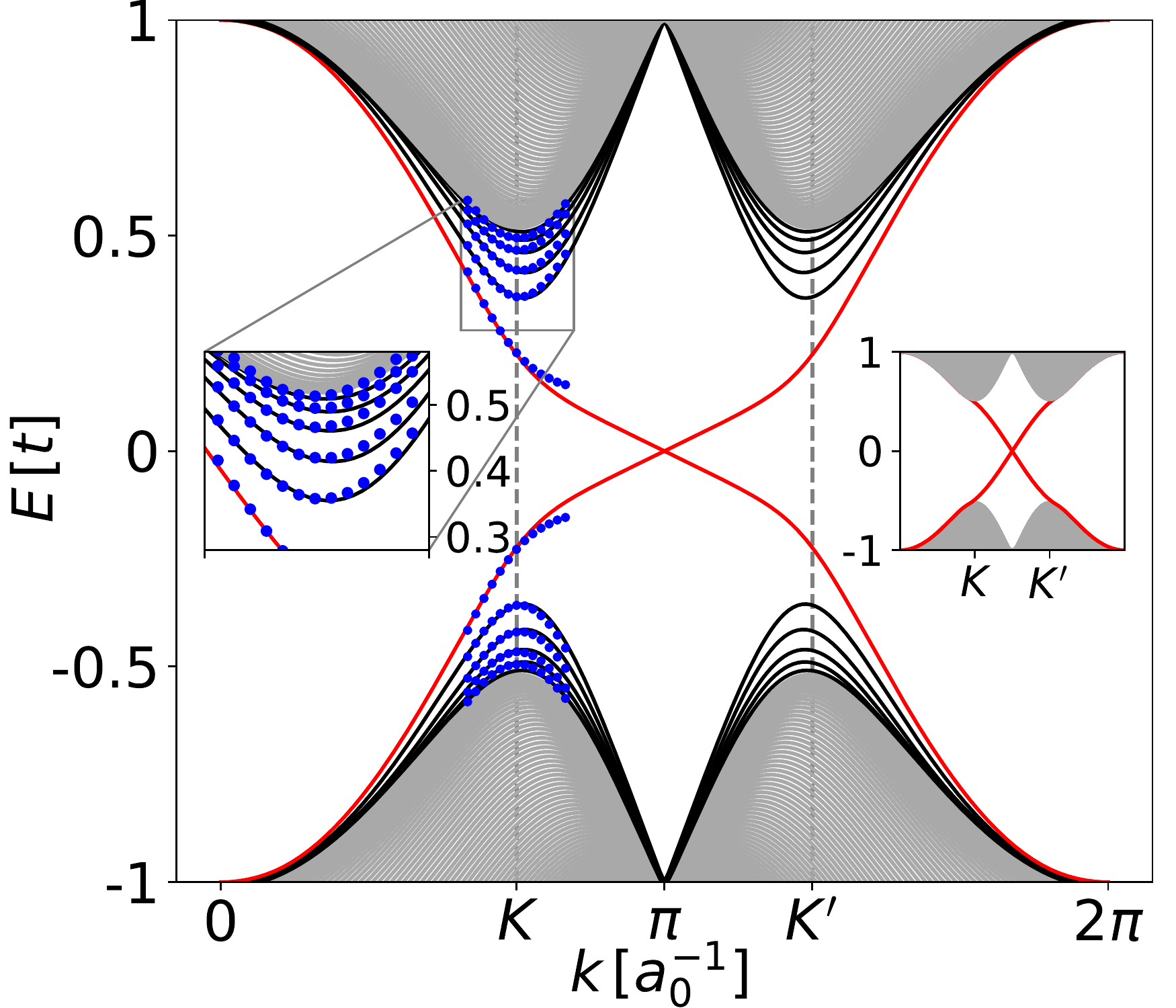}
    \caption{\label{fig_Bandszigzag} Band structure of a zigzag nanoribbon of width $L_y = 37.2$ nm ($150$ rows), 
    with modulation of the SOC at the boundary. We set $\lambda_\text{i}=0.1t$, $\lambda_\text{e}=-0.05 t$, and $\ell=12 a_0$.
     The gray, black, and red lines are the tight-binding results, describing  bulk  states,  VP  states,  and topological states, respectively. 
     The analytical results from Eq.~\eqref{eq_zigzagFinalEq} are represented as blue dots. The right inset shows the case of homogeneous SOC,
     with $\lambda=0.1t$, corresponding to a gap $\Delta=3\sqrt{3}\lambda\approx 0.52t$.
     In both cases, two topological modes (red lines) cross within the gap. The modulation of the SOC results in additional massive edge states, 
     popping up under the conduction band and above the valence band (black lines).}
\end{figure}

Before discussing the solutions of Eq.~\eqref{eq_zigzagFinalEq}, we notice that this
equation is invariant under the following transformations:
%
%
%%%%%%%%%
\begin{align*}
&\tau,s,k_x \rightarrow -\tau,-s,-k_x\, , \\
& E,k_x,\alpha \rightarrow -E,-k_x,-\alpha  \, . 
%& E,\tau,s,\alpha \rightarrow -E,-\tau,-s,-\alpha \, .
\end{align*}
%%%%%%%%%%%%
%
%
The first invariance is just the consequence of the time-reversal symmetry 
of the Hamiltonian~\eqref{LWAHamiltonian}: the edge states occur in pairs of 
counterpropagating modes with opposite spins, residing on opposite valleys.
The second invariance implies that the edge state spectrum at a $B$-type edge 
can be obtained from the spectrum at an $A$-type edge by simply reversing energy 
and wave vector. In order to reproduce the full spectrum of edge states (including degeneracies) 
of a wide zigzag nanoribbon, which has one edge of  $A$ sites and one edge of $B$ sites, we need 
to take the solutions of Eq.~\eqref{eq_zigzagFinalEq} with $\alpha=+1$ and with $\alpha=-1$. 
These solutions are shown in Fig.~\ref{fig_Bandszigzag} as blue dots on top of the numerical 
results discussed in the next sections (gray, black, and red lines). For clarity, 
we only include the states at one valley, the states at the other valley follow by symmetry.
%
%
%%%%%%%%%
%%%%%%%%%%%%
%
%

We observe that the continuum approach faithfully reproduces all the main 
features of the spectrum close to the Dirac points. Within the gap, there 
exist two topological bands with approximately linear dispersion, and ten 
massive VP modes (for the given values of the parameters), in agreement with 
the results of the numerical approach discussed in Sec.~\ref{numerics} below. 
All these levels are doubly degenerate if one considers a system with two edges.  
Figure~\ref{fig_Bandszigzag} shows the agreement between analytical and numerical results. 
For wave vectors in the interval between the two Dirac points, the agreement is less 
satisfactory, which we attribute to the fact that the coupling between the valleys, 
neglected in the continuum approach, plays an important role at these wave vectors. 
This is especially evident for the topological bands.
Another source of discrepancy stems from neglecting higher order terms in 
momentum in the LWA.

%%%%
\subsection{Spectral properties of armchair ribbons}
\label{analyticsarmchair}
%%%%

Let us now turn to the case of a semi-infinite system with an armchair edge along the $y$-direction 
[see Fig.~\ref{fig_system}(b)]. The wave function can be written as  
%
%
%%%%%%%%%
\begin{align}\label{wfarmchair}
\Psi(x,y)=e^{ik_y y}\psi(x)\, ,
\end{align}
%%%%%%%%%
%
%
with $\psi^T(x)=(v_A ,v_B)$. Then the Dirac equation reduces to
%
%
%%%%%%%%%
\begin{align}\label{diracarmchair}
\left[ - i\tau \sigma_x \partial_x + \sigma_y k_y +  s \tau \Delta(y) \sigma_z\right]\psi=E\psi \,,
\end{align}
%%%%%%%%%
%
%
and its solutions can be written as
%
%
%%%%%%%%%
\begin{subequations}
\begin{align}
v_A &= u^{\kappa_\text{i}/2}(1-u)^{\kappa_\text{e}/2} (\tilde w_+ + \tilde w_-)\,,\\
v_B &= i u^{\kappa_\text{i}/2}(1-u)^{\kappa_\text{e}/2} (\tilde w_+ - \tilde w_-)\,, 
\end{align} 
\end{subequations}
%%%%%%%%%%%%
%
%
where $u=\frac{1}{2}[1-\tanh(x-x_0)]$. The functions $\tilde w_\pm$ are,
again, solutions of a hypergeometric equation, and are given  by
%
%
%%%%%%%%%
\begin{align}
\tilde w_\pm(u) &= d_\pm \, F[ \bar \kappa \mp s\Delta_0, \bar \kappa \pm s \Delta_0 +1 ; \kappa_\text{i} +1; u], 
\end{align}
with the relative factor, fixed by the Dirac equation, given by 
\begin{align}
\frac{d_-}{d_+} = \frac{\tau (s \Delta_\text{i} - \kappa_\text{i})}{E+k_y}\, .
\label{armchairprefactor}
\end{align}
%%%%%%%%%%%%
%
%
Notice that the dependence on the valley index only appears in the 
prefactor.
%
%
%%%%%%%%%
%\begin{subequations}
%\begin{eqnarray}
%w_+ &= &(E+k_y) \times  \\
%&& F[ \bar \kappa - s\Delta_0, \bar \kappa + s \Delta_0 +1 ; \kappa_\text{i} +1; u],\nonumber \\
%w_- &=&\tau (s \Delta_\text{i} - \kappa_\text{i}) \times  \\
%&& F[\bar \kappa -s\Delta_0+1, \bar \kappa +s\Delta_0  ; \kappa_\text{i} +1; u] \nonumber.
%\end{eqnarray}
%\end{subequations}
%%%%%%%%%%%%
%
%

In the armchair case, the boundary condition involves both valleys \cite{Brey_2006,CastroNeto_2009}
and reads
%
%
%%%%%%%%%
\begin{align}
\left[ v_{\alpha,\tau=+1,s} + v_{\alpha,\tau=-1,s}\right]_{x=0}=0 \, , \quad s=\pm 1\,,
\end{align}
%%%%%%%%%%%%
%
%
with $\alpha=A,B$. In terms of the $w_\pm$ we find
%
%
%%%%%%%%%
\begin{align}
\left[ \tilde w_{\pm,\tau=+1} + \tilde w_{\pm,\tau=-1}\right]_{u=u_0}=0 \, , \quad s=\pm 1\, .
\end{align}
%%%%%%%%%%%%
%
%
From Eq.~\eqref{armchairprefactor} we see that either $\tilde w_{-,\tau=-1}=-\tilde w_{-,\tau=1}$ 
or $\tilde w_{+,\tau=-1}=- \tilde w_{+,\tau=1}$,
therefore the boundary condition takes the simple form 
%
%
%%%%%%%%%
\begin{align} \label{eq_armchairFinalEq}
(E \pm k_y)
F[\bar \kappa  \mp s\Delta_0, \bar \kappa  \pm  s \Delta_0 +1 ; \kappa_\text{i}  +1; u_0]=0\, .
\end{align}
%%%%%%%%%%%%
%
%
Here, $u_0=\frac{1}{2}(1+\tanh x_0)$. In the armchair system, the shift required 
to have the correct value of the SOC on the first line of carbon atoms is $x_0=a_0/2$. 

\begin{figure}
    \centering
    \includegraphics[width=\linewidth]{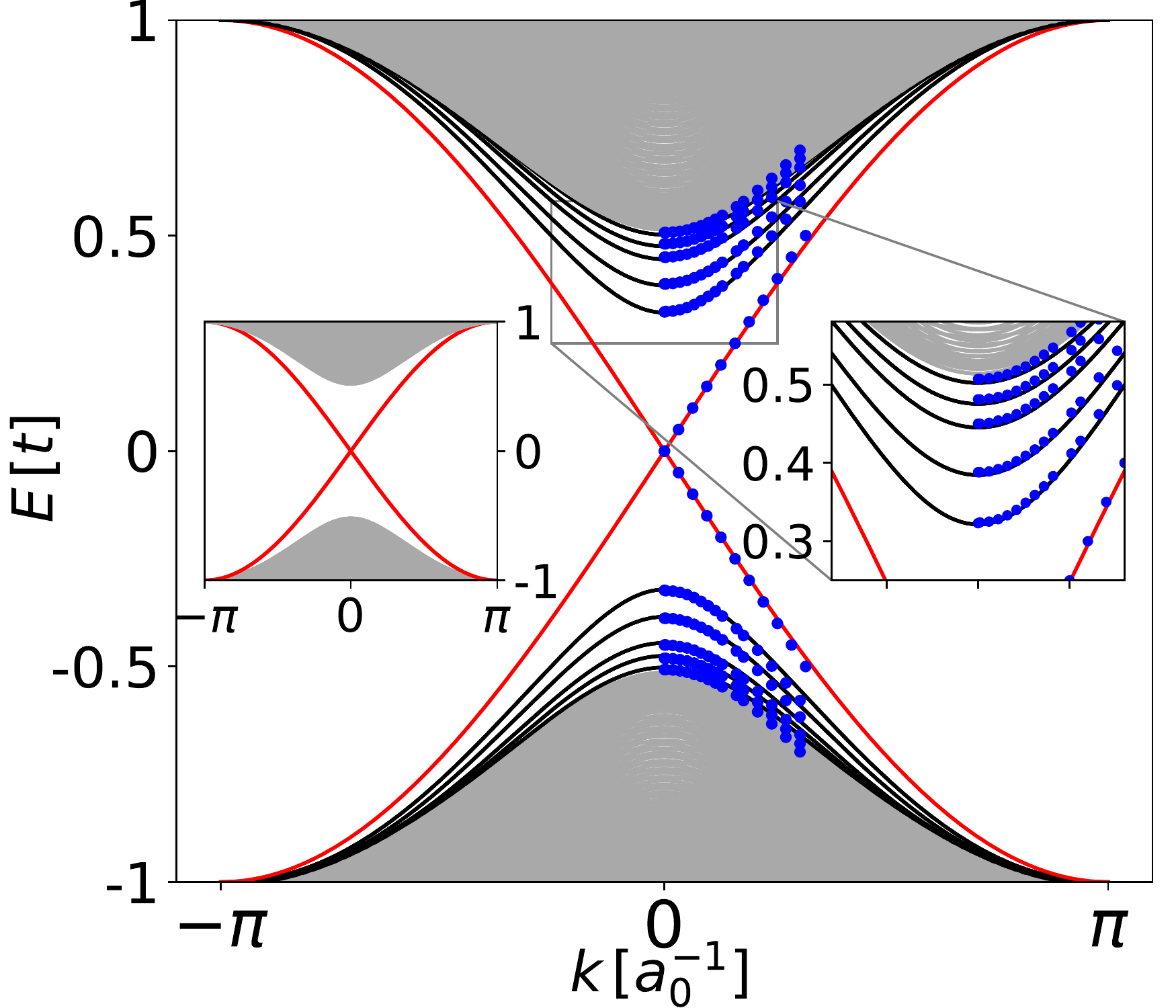}
    \caption{\label{fig_BandsArmchair}Band structure for an armchair nanoribbon of width $L_x = 32.5$ nm  ($128$ rows), 
    with SOC modulation near the boundary. We set $\lambda_\text{i}=0.1t$, $\lambda_\text{e}=-0.05 t$, and $\ell=12 a_0$. 
    The gray, black, and red lines are the tight-binding results, describing respectively bulk states, VP states, 
    and topological states. The blue dots are the solutions of Eq.~\eqref{eq_armchairFinalEq}. 
    Close to the $\mathbf{K}$ ($\mathbf{K'}$) point (in the armchair case, projected to $k=0$), the agreement is excellent. 
    The right inset shows the case of homogeneous SOC with $\lambda=0.1t$, corresponding to a gap $\Delta=3\sqrt{3}\lambda\approx 0.52t$.}
\end{figure}

The solutions to Eq.~\eqref{eq_armchairFinalEq} are shown in Fig.~\ref{fig_BandsArmchair} as blue dots. 
As in the zigzag case, in a nanoribbon all levels are doubly degenerate, corresponding to states on both edges. 
The obvious solutions $E= \mp k_y$ describe  two counterpropagating linearly dispersing topological bands. 
These modes only exist if $s=\mp 1$, respectively, otherwise one gets the trivial solution $\tilde{w}_+=\tilde{w}_-=0$. 
The allowed value of $s$ guarantees that the corresponding wave function is normalizable. 
This is a manifestation of the spin-momentum locking. Remarkably, in contrast to the zigzag case, 
here the group velocity of the topological modes is equal to $v_F$ and does not depend on SOC. Moreover, 
we observe that, for $E\neq \pm k_y$, the boundary condition depends on $k_y$ and $E$ only through the combination 
$k_y^2-E^2$. Therefore we find solutions corresponding to VP states, whose dispersion has the form
%
%
%%%%%%%%%
\begin{align}
   E_n = \pm \sqrt{M^2_n+k^2_y}\, , \quad n=1,\dots,N_\text{max} \, ,
\end{align}
%%%%%%%%%%%%
%
%
where the effective masses $M_n$  and the number of massive states  $N_\text{max}$  depend on the parameter values. 
In Fig.~\ref{fig_BandsArmchair}, we compare the analytical results with the numerical ones,
finding an excellent agreement. 
Since in this case the continuum approach incorporates the coupling
between valleys in the boundary condition, it is not surprising that
the agreement is better than in the zigzag case. 
%
%
%%%%%%%%%
%%%%%%%%%%%%
%
%

%%%%%%%%%%%%%%%%%%%%%%
\section{Numerical tight-binding model}\label{numerics}
%%%%%%%%%%%%%%%%%%%%%%
In order to be able to go beyond the low-energy approximation, as well as to study the transport properties, 
we now move on to a fully numerical approach within a tight-binding formalism, which we implement using Kwant~\cite{Groth_2014}.  
Contrary to the analytical calculation, we will here consider a finite size system, comprising of a scattering region with two edges, 
along which edge states can propagate, and a source and a drain lead, which are seamlessly coupled to the scattering region. We will take rather large 
ribbons, but still of experimentally relevant sizes, with length $L \approx 60$~nm and width $W  \approx 35$~nm. 
For this width, which largely exceeds the modulation length $\ell = 12\,a_0 \approx 3$~nm, 
the electronic states located on opposite edges do not overlap. Hence, the edges are independent, 
and the numerical results can readily be compared to the analytical solution for a semi-infinite system.
For the local 
%%%%
\subsection{The model and its parameters}
%%%%
Within a tight-binding formalism, the Hamiltonian for graphene with intrinsic SOC reads~\cite{Kane_2005A,Kane_2005B}:
%
%
%%%%%%%%%%%%
\begin{align}\label{Ham}
\mathcal{H}=-t\!\!\sum_{\substack{\langle n,m \rangle\\ s}}c_{n s}^\dag c_{m s} + i \lambda \!\!\!\!\! \sum_{\substack{\langle\langle n,m \rangle\rangle\\ s s'}} \nu_{nm} (s_z)_{s s'} c_{n s}^\dag c_{m s'}\, ,
\end{align}
%%%%%%%%%%%%
%
%
where $c_{n s}^\dag$ ($c_{n s}$) creates (annihilates) an electron with spin $s$ on the site $n$, and 
the symbol $\langle \ldots \rangle$ ($\langle\langle \ldots \rangle\rangle$) indicates sum over nearest (next nearest) neighbour sites. 
In Eq.~\eqref{Ham}, the sign $\nu_{nm}=\pm1$ depends on the orientation of the  next nearest neighbour hopping: it is positive (negative) 
for electron making a left (right) turn to the next nearest neighbour carbon atom. The hopping parameter is $t$, and $\lambda$ is 
the intrinsic SOC parameter, which is related to the gap size as $\Delta = 3 \sqrt{3} \lambda$.

%As for the continuous case, 
We consider the following space modulation of the intrinsic SOC along the coordinate corresponding to the lateral width of the graphene nanoribbon:
%
%
%%%%%%%%%%%%%
\begin{align}\label{lambda_modulation}
\lambda (\zeta) &= \frac{ \lambda_\text{i} + \lambda_\text{e}}{2}  \\
&+ \frac{\lambda_\text{i} - \lambda_\text{e}}{2} \Big[ \tanh \left( \frac{\zeta}{\ell } \right) -  \tanh \left( \frac{\zeta-L_\zeta}{\ell}  \right) -1 \Big] \,, \nonumber
\end{align}
%%%%%%%%%%%%
%
%
where $L_\zeta$ is the width of the ribbon in the $\zeta$-direction, and $\lambda_\text{i(e)}$ is the value of the SOC 
in the internal (external) region of the ribbon, respectively.  Throughout this paper, we use $\lambda_\text{i}=0.1t$ and $\lambda_\text{e}=-0.05t$. 
The length scale $\ell$ characterizes the size of the spatial region over which the variation of the intrinsic SOC takes place. 
This has to be compared with the three natural length scales present in the system: the lattice constant $a_0$, the length 
scale associated to the SOC gap $\xi=\hbar v_\text{F}/\Delta$, and the width of the ribbon $L_\zeta$. 
In order to get VP states, one has to assure $\ell \gtrsim \xi$. Moreover, to resolve the smoothness of the SOC modulation one needs $\ell \gg  a_0$, and the two edges are independent for $L_\zeta \gg \ell$.
In App.~\ref{app_parameters} we provide some details on the relation between the parameter values in the LWA and in the tight-binding model.

%%%%
\subsection{Spectral properties}
%%%%
In this section we investigate the spectral properties of graphene nanoribbons: the band structure and the local density of states (LDOS). The band structure for homogeneous SOC are shown in the insets of Figs.~\ref{fig_Bandszigzag} and~\ref{fig_BandsArmchair}. 
\comment{We measure energy in terms of $t$, and we take $\lambda=0.1 t$, which corresponds to a gap of size $\Delta = 3 \sqrt{3} \lambda \approx 0.52 t$. }
Without SOC modulation, each edge of the system hosts two topological states with linear dispersion, as shown in the insets in red. 
The bulk states, in gray, form the conductance band (CB) and the valence band (VB).

For the calculation of the LDOS, Kwant first calculates the available modes in the semi-infinite source lead. These modes are then projected onto the scattering matrix of the scattering region. 

%%%%
\subsubsection{Zigzag ribbons}

%%%%
%
%
%%%%%%%%%%%%
\begin{figure}[!t]
    \centering
    \includegraphics[width=\linewidth]{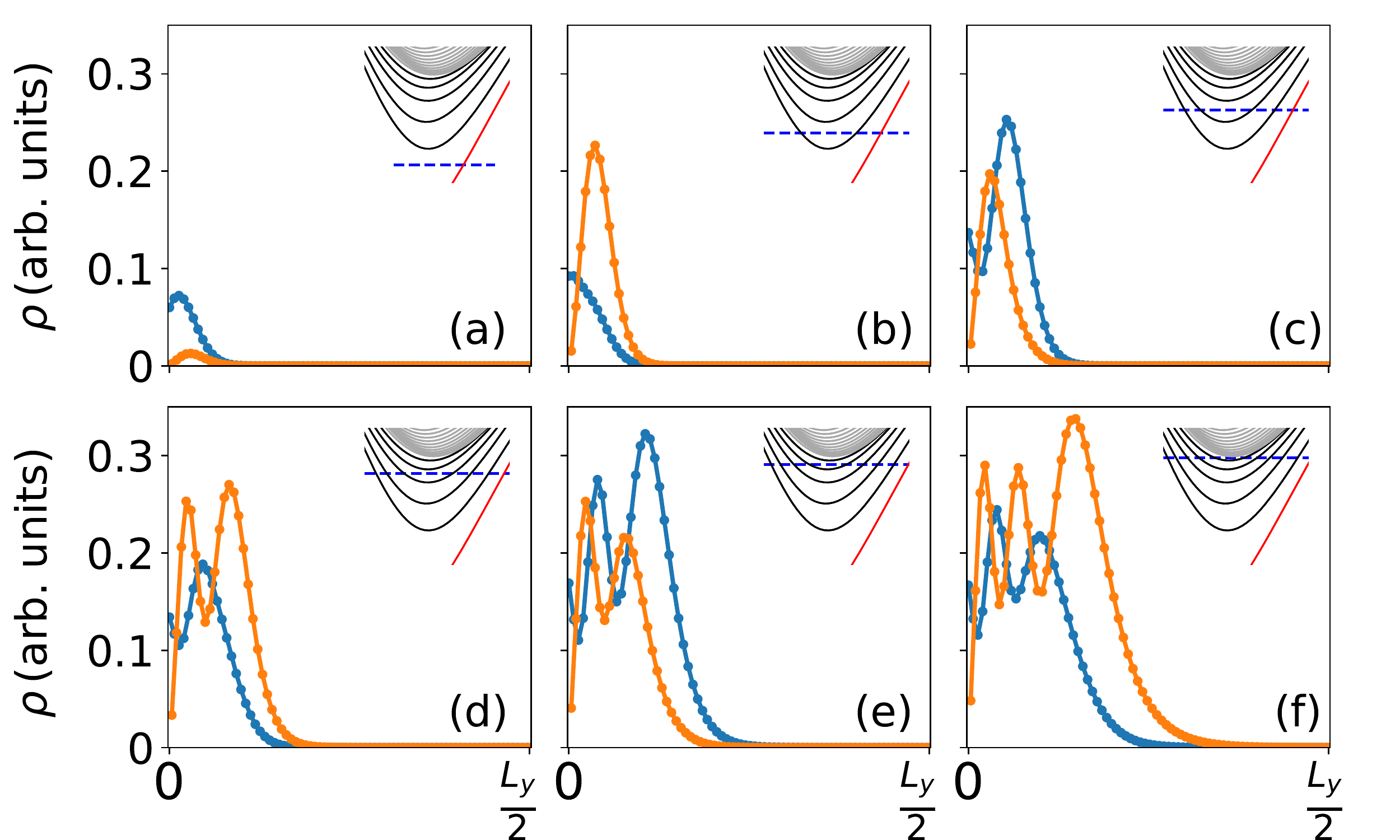}
    \caption{\label{fig_Ldoszigzag}Local density of states at one edge of a zigzag nanoribbon, for $A$ (orange) and $B$ (blue) sublattices, 
    summed over the spin. On the opposite edge, $A$ and $B$ sublattices should be exchanged. In the inset, the energy 
    at which the LDOS is evaluated is indicated by the dashed-blue line in the band structure. 
    }
\end{figure}
%%%%%%%%%%%%
%
In the zigzag case, we consider a ribbon of width $W = L_y = 37.2 $~nm,  corresponding to $150$ rows. Figure~\ref{fig_Bandszigzag}
shows that, due to the suppression of the SOC near the edges, VP bands are pulled out of the CB and the VB, in symmetric fashion. 
Near the $\mathbf{K}$ and $\mathbf{K}'$ points, these VP modes push away the topological modes, 
whose group velocity is thus strongly affected. 
This feature can be rationalized by observing that, from Eq.~(\ref{eq_zigzagFinalEq}), one can see that 
the group velocity of the topological modes close to the Dirac points depends strongly on
the gap parameters $\Delta_\text{i/e}$. 
\comment{It can indeed be seen from the analytical expression of Eq.~(\ref{eq_zigzagFinalEq}), when solved for $k_x=0$ giving $E_0$ 
and then expanded around $E_0$, that the group velocity depends on $\Delta$, the gap parameter.}

Near the boundary, as the SOC gets weaker, the effective gap becomes smaller. 
Looking at the VP modes under the CB, we observe that  the first (lowest in energy) VP mode is the one which lies closest 
to the edge, and has the smallest decay length. Each consecutive VP mode has a longer decay length and
extends deeper in the bulk, and therefore ``sees" a slightly larger effective gap.
\comment{Each of the consecutive VP modes, lies slightly farther away from the edge and has a slightly larger effective gap.}

In Fig.~\ref{fig_Ldoszigzag} the LDOS of the edge states is plotted at different energies. 
In Fig.~\ref{fig_Ldoszigzag}(a) there is only the topological mode, in \ref{fig_Ldoszigzag}(b) there is additionally one VP mode, 
in \ref{fig_Ldoszigzag}(c) two VP modes etc. In the zigzag case we observe that there are zones on the lattice with predominant 
$A$ or $B$ contributions to the LDOS. Moreover, for each VP mode that is added, the predominance changes sublattice. 

%%%%
\subsubsection{Armchair ribbons}
%%%%
%
%
%%%%%%%%%%%%
\begin{figure}[!t]
    \centering
    \includegraphics[width=\linewidth]{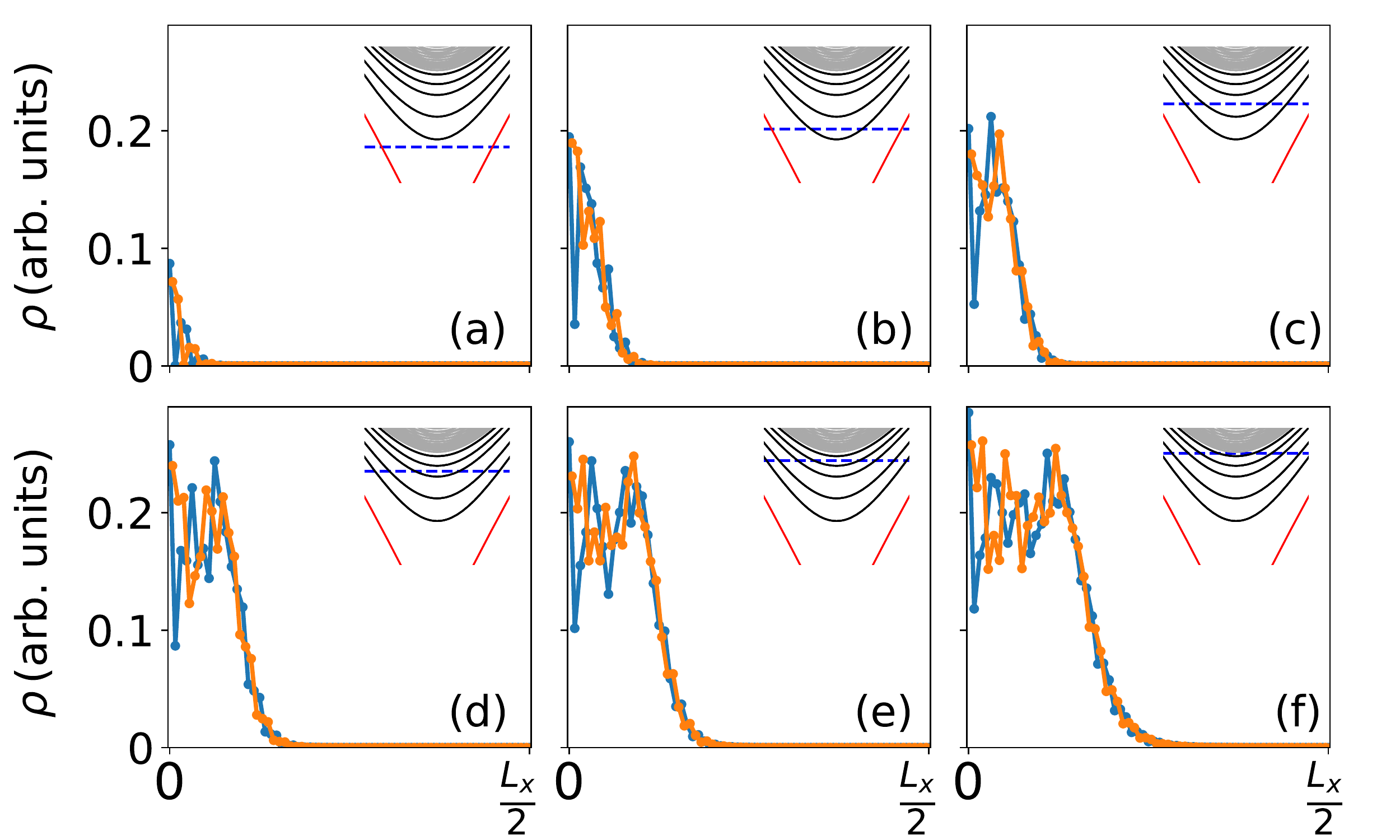}
    \caption{\label{fig_Ldosarmchair}Local density of states at one edge of an armchair nanoribbon, for A (orange) and B (blue) sublattice, summed over the spin, with the corresponding energy indicated in the inset as a dashed-blue line in the band structure.}
\end{figure}
%%%%%%%%%%%%
%
%
In the armchair case, we consider a ribbon of width $W=L_x=32.5$~nm, corresponding to $128$ unit cells. 
The band structure is shown in Fig.~\ref{fig_BandsArmchair}.
The two Dirac points are projected onto the same point $k=0$. In the absence of  SOC, an armchair ribbon is metallic or semiconducting, 
depending on its exact width \cite{CastroNeto_2009}. In the case studied here, the ribbon is wide enough that the semiconducting gap, 
which is of order of $\hbar v_F/L_x$, is exceedingly small compared to all other energy scales, and can be ignored. 
In the presence of SOC, the gap size is then determined by the SOC strength, and there are always two topological modes crossing the gap. 
With the SOC modulation, new massive bands appear under the CB and above the VB. However, in contrast to the zigzag case,
the dispersion of the topological bands is not affected by the appearance of these new levels, and their group velocity is not modified. 
This is consistent with the analytical solution of Eq.~(\ref{eq_armchairFinalEq}), which gives a linear dispersion with slope $v_F$, independent 
of the values of the gap parameters. As shown in Fig.~\ref{fig_BandsArmchair}, the agreement between the analytical and numerical results 
for all edge states is excellent.

In the LDOS for an armchair ribbon we see there are no privileged $A$ or $B$ sublattice zones | Fig.~\ref{fig_Ldosarmchair}. 
It is harder in this case to say where on the lattice the different VP states lie. Also in this case, we observe the expansion of 
the area containing the edge states, as one consecutively adds subbands.

%%%%
\section{Transport properties}\label{transport}
%%%%
In this section we will investigate if two-terminal conductance measurements on nanoribbons with SOC 
modulation near the edges could reveal the presence of VP states. Being massive, VP states are not protected against backscattering due to disorder.
%subject to disorder. 
We will therefore add Anderson disorder to the model, namely random spin-independent onsite energies with uniform probability
distribution in the interval $\varepsilon_n \in [-U_0/2,\,U_0/2]$. The disorder Hamiltonian can be written as
%
%
%%%%%%%%%%%%
\begin{align}
    H_\text{D} = \sum_{n s} \varepsilon_n c_{n s}^\dag c_{n s}\, ,
\end{align}
%%%%%%%%%%%%
%
%
where the sum runs over the entire system. The topological edge states will not be sensitive to this disorder, 
but the VP states will. How much the disorder weakens the VP states depends not only on the disorder strength, 
but also on the system size. Here, we consider a sample of size $L \times W \approx 60 \times 35$~nm$^2$.  

%%%%
\subsection{In-gap conductance}
\begin{figure}
    \centering
    \includegraphics[width=\linewidth]{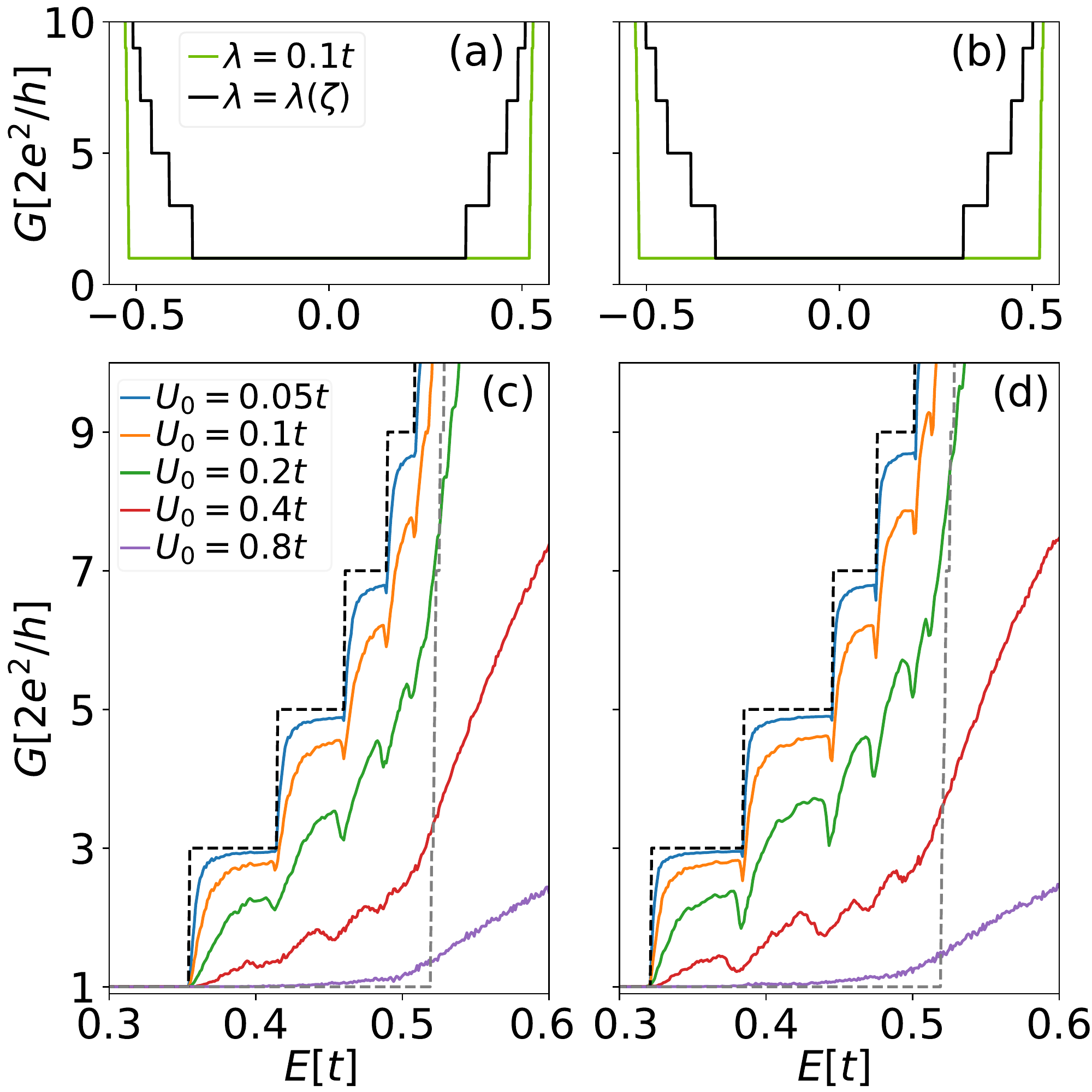}
    \caption{\label{fig_G}Conductance for a clean ribbon without (grey) and with (black) modulation of the SOC interaction for a zigzag (a) and armchair (b) ribbon. In the case with modulation $\lambda_\text{i}=0.1 t$ and $\lambda_\text{e}=-0.05 t$. The lower panels (c)-(d) are a zoom around the CB edge, where the VP modes are, for several disorder strengths. Within the range $U_0 \in [0.05,  0.4]$ clear conductance minima are observed just before the opening of a new VP mode.}
\end{figure}

%%%%
The numerical results for the two-terminal conductance are shown in Fig.~\ref{fig_G}.
As one can expect for a system presenting in-gap states, the conductance never decreases to zero. 
Here, we have two counterpropagating  doubly-degenerate topological edge states in the gap, so the minimal in-gap conductance is $2e^2/h$. 
In the absence of SOC modulation, this is the only in-gap contribution to the conductance. With SOC modulation near the edges, 
we observe clear steps at the energies at which new VP modes open, see upper panels of Fig.~\ref{fig_G}. Both in the zigzag as well as 
in the armchair cases, these steps are symmetric around the gap center $E=0$. 

In the presence of disorder, the conductance due to the 
VP modes is suppressed, because those modes are sensitive to backscattering. For strong disorder, the conductance reduces to $2 e^2/h$, 
below which it can not descend because the topological modes are not sensitive to Anderson disorder, which does not break time-reversal symmetry. 
However, in this case the conductance in the CB and the VB is also significantly reduced due to the disorder in the ribbon.
Remarkably, within a certain range of disorder strength, we observe dips in the conductance at the step edges. 
This reminds of the physics of a quasi-1D quantum wire containing an impurity with an attractive potential, where 
the conducting modes couple to a quasi-bound state at the impurity~\cite{Bagwell_1990}. 
%
%
%%%%%%%%%%%%
%%%%%%%%%%%%
%
%

%%%%
\subsection{Dip behavior near the steps}
%%%%

In the lower panels of Fig.~\ref{fig_G}, we observe a minimum at each step in the conductance curve of the disordered system. 
Notice that each line represents the conductance averaged over 100 disorder configurations. Plotting single disorder configurations 
[c.f. App.~\ref{app_ConductanceDips}], one observes random fluctuations at the plateaus, which average out over many disorder configurations, 
as well as a dip at the step edges, which most configurations have in common, and which therefore remains in the averaged conductance. 
Such dips in the conductance at the opening of new subbands were discussed in detail in a paper by Bagwell~\cite{Bagwell_1990,Tekman_1993}, 
in which he shows how propagating modes in a narrow wire with parabolic confinement couple to the zero energy quasi-bound states of delta shaped, 
negative potential impurities. 
Similar dips were also explored in the context of quantum Hall states~\cite{Palacios_1993,haug1993edge}.
In the case presented here, in the energy gap we have to deal with a quasi-1D subsystem near the edge, 
with triangular confinement potential. The quasi-bound states can be hosted in local energy minima, that appear due to the random energy landscape. 
In order to verify that we are indeed dealing with this physical phenomenon, we have simulated a clean sample, containing one Gaussian-shaped impurity 
at each edge. We investigated the minimal requirements for observing dips in the conductance curves right at the onset of the steps. The simulations with single impurities clearly demonstrate that
the dips come from the coupling of the propagating states to quasi-bound states lying at randomly distributed energy minima on the lattice. This can be observed in strongly confined quasi-1D systems with an attractive impurity (negative potential). 
The precise properties of the dip in the conductance depend on the shape and strength of the impurity, and other system details. Additional information is given in App.~\ref{app_ConductanceDips}.
 
Although the situation near the edges in the system presented here differs from that investigated in earlier works, many of the physical arguments still hold~\cite{Chu_1989aa,Bagwell_1990}. Because of the random energy landscape, attractive sites or regions on the lattice may result in the appearance of quasi-bound states lying just under each VP subband. As we can observe, and as is normally the case, the evanescent state under the lowest VP subband is a true bound state, as it does not couple to the topological state. We therefore observe no dip before the first step, i.e. conductance never decreases below $2 e^2/h$. 
%As discussed by Bagwell, the conductance value at the energy band minimum (at the step), is not affected in the case for weak delta shaped impurities, because at these energies the normal modes decouple from the evanescent modes and ballistic transport is recovered. This is generally the case for the class of isotropic, or $s$-like impurities, to which delta shaped impurities belong~\cite{Chu_1989aa}. Because we have random on-site Anderson disorder, we also observe this feature for weak enough disorder. For stronger disorder, because we then couple multiple VP modes, this no longer holds. 
We also observe the usual renormalization of the energy gap, which slightly shifts the energies of the onsets of the steps, and the value of the CB opening~\cite{Groth_2009aa, Jiang_2009aa, Wu_2016aa}. The binding energy, defined as the difference in energy between the step edge and the dip minimum, is therefore hard to quantify. 
However, we do observe that, as disorder gets stronger, the dips become deeper, which is in line with earlier observations. In wires with a parabolic confinement potential and single attractive impurities, the interaction strength between the impurity and the various available subbands depends on the lateral position of the impurity in the wire. For our triangular confinement near the edge there is no symmetry around any center, however, we know where at the edge each of the modes lies, and if it can interact with the impurity. In the implementation of our system with one single impurity at each edge, we clearly observe how the impurity interacts with consecutive edge states as we move it away from the edge towards the bulk of the sample, as discussed above. 
A more in depth investigation is beyond the scope of this work, but will be the subject of a future investigation.

%%%%%%%%%%%%%%%%%%%%%%
\section{Conclusion and outlook}
\label{conclusions}
In this work, we have investigated the appearance of Volkov-Pankratov edge states in topological graphene nanoribbons of zigzag and armchair type. 
In the presence of intrinsic SOC which is smoothly suppressed near the edges, the well-known QSH edge states are accompanied by multiple massive 
VP states. We have demonstrated their existence by means of two complementary methods, the exact analytical solution of the low-energy effective Dirac equation,  
and the numerical tight-binding approach, finding good agreement between the two. 

Transport simulations show how the VP modes contribute to transport, also in the presence of disorder. 
We observe dips in the conductance at the onset of each VP band, which are due to the coupling of the propagating VP states
to evanescent modes present in the random energy landscape. At sufficiently strong disorder, the VP states are entirely suppressed, 
and cease to contribute to transport. 

Our results can be relevant to experimental systems in which the intrinsic SOC in graphene is enhanced by 
one of the methods mentioned in the Introduction. Both the deposition of adatoms, as well as the proximitization 
with a TMD layer, would likely give rise to a inhomogeneous intrinsic SOC, especially at the edges of the system.

In addition to a possible implementation in graphene or “post-graphene”  materials, the presence of these VP states accompanying the 
topological modes could be achieved in systems of ultracold atoms in optical lattices. 
The advantage of a realization within this platform is related to the flexibility to control the parameters separately across a large range  compared to condensed-matter systems, where the system parameters are generally fixed by the material properties and by the sample geometry~\cite{Bloch_2008}. 

\comment{The time-reversal symmetric Kane-Mele model for the QSH effect~\cite{Kane_2005B,Kane_2005A} can be thought of 
as a double copy of the Haldane model~\cite{Haldane_1988}, in which time-reversal symmetry is broken.
As  both the honeycomb lattice and the Haldane model have already been realized in the context of ultracold atoms~\cite{Tarruell_2012,Jotzu_2014}, 
the implementation of the Kane-Mele model is achievable with the state-of-the-art technology.} 
The time-reversal symmetric Kane-Mele model for the QSH effect~\cite{Kane_2005B,Kane_2005A} can be thought of as a double copy of the Haldane model~\cite{Haldane_1988} in which time-reversal symmetry is broken.
The implementation of the Kane-Mele model could be achieved within the state-of-the-art technology for ultracold atoms;  
the honeycomb lattice and the Haldane model have been already realized in this context~\cite{Tarruell_2012,Jotzu_2014}.
In practice, the Kane-Mele model could be implemented by using an internal atomic state as a spin degree of freedom. For each spin, 
the same scheme as for the Haldane model could be used to implement the second-next-neighbour hopping, see Ref.~\cite{Goldman_2010,Bercioux_2011,Goldman_2011}. 
This system would then correspond to two copies of the Haldane model~\cite{Jotzu_2014}. 
Contrary to the implementation in Ref.~\cite{Goldman_2016},  we propose to realize a system with a homogeneous intrinsic SOC and soft-boundary conditions, corresponding to an inhomogeneous onsite energy profile due to the confining potential of the atomic trap. Similar to the results presented in
 this work, this scheme gives rise to a set of VP states accompanying the topologically protected one, but with energy symmetry breaking. 
 A similar approach was proposed for the case of the quantum Hall edge states~\cite{Buchhold_2012}. 
Other aspects of the implementation of this model for ultracold atoms require further investigation.

%%%%%%%%%%%%%%%%%%%%%%
%%%%%%%%%%%%%%%
\section*{Acknowledgements}
Discussions with N. Goldman, M. Pelc, and V. Golovach are acknowledged. 
The work of TB and DB supported by the Spanish Ministerio de Ciencia, Innovation y Universidades (MICINN) through the project FIS2017-82804-P, and by the Transnational Common Laboratory \emph{Quantum-ChemPhys}. MRC acknowledges funding from the Spanish Government through project MAT2017-88377-C2-2-R (AEI/FEDER) and the Generalitat Valenciana through grant Cidegent2018004.

%%%%%%%%%
\appendix
%%%%%%%%%%%%%%

\section{Solution of the Dirac equation with inhomogeneous SOC}
\label{analytic_solution}

In order to make this paper self-consistent, in this appendix we provide the details of the 
exact solution of the Dirac equation in the presence of an inhomogeneous SOC 
with hyperbolic tangent profile. The analysis follows Refs.~\cite{Tchoumakov_2017,LandauQM}. 

%%%%%%%%
\begin{widetext}
\subsection{Zigzag case}

We consider first the case of a semi-infinite system with a zigzag edge along the $x$-direction,
see Fig.~\ref{fig_system}(a). After factorization of a plane wave in the $x$-direction with wave
vector $k_x$, the Dirac equation reads
%
%
%%%%%%%%%%%%
\begin{align}\label{Diraczz}
\left\{\tau \sigma_xk_x -i \sigma_y \frac{d}{dy} +\sigma_z s \tau  
\left[ \bar \Delta+ \Delta_0\tanh(y-y_0)\right] \right\} \psi(y) =E \psi(y)\, ,
\end{align}
%%%%%%%%%%%%
%
%
where we set $\hbar=1$, and measure energies in units of $v_F/\ell$, 
lengths in units of $\ell$, and wave vectors in units of $\ell^{-1}$. 
We look for solution on the half-line $y\geq 0$.
Squaring Eq.~\eqref{Diraczz}, we obtain
%
%
%%%%%%%%%%%%
\begin{align}
\left[ \frac{d^2}{dy^2} + (E^2-k^2_x-\bar \Delta^2- \Delta^2_0) + 
\frac{\Delta_0(\Delta_0- s\tau \sigma_x)}{\cosh^2(y-y_0)} -
2 \bar \Delta \Delta_0 \tanh(y-y_0) \right] \psi(y) =0\, .
\end{align}
%%%%%%%%%%%%
%
%
Next, we express $\psi$ as
%
%
%%%%%%%%%%%%
$$
\psi =\left( \begin{array}{c}
v_A\\v_B
\end{array}\right) = \phi_+ |+\rangle + \phi_- |-\rangle, 
\qquad v_{A/B}=\frac{1}{\sqrt{2}} \left( \phi_+ \pm \phi_-\right)\, ,
$$
%%%%%%%%%%%%
%
%
where $|\pm \rangle$ are the eigenvectors of $\sigma_x$, with respective eigenvalue $\pm 1$.
By performing the change of variable  
$$
u=\frac{1}{2}\left[1-\tanh(y-y_0) \right]\, ,
$$ 
Eq.~\eqref{Dirac2} can be rewritten as 
%
%
%%%%%%%%%%%%
\begin{align}
\left[ u(1-u) \frac{d}{du} u(1-u) \frac{d}{du} -
\frac{1}{4} (\kappa_\text{i}^2(1-u) +\kappa_\text{e}^2u) + 
(\Delta^2_0-rs\tau\Delta_0)u(1-u)  \right] \phi_r =0,\quad r=\pm \, ,
\label{Dirac2}
\end{align}
%%%%%%%%%%%%
%
%
where we use the notation
%
%
%%%%%%%%%%%%
$$
\kappa_\text{i/e} = \sqrt{k_x^2+\Delta^2_\text{i/e}-E^2}\, , \quad \bar \kappa = 
\frac{\kappa_\text{i}+\kappa_\text{e}}{2}\, .
$$
%%%%%%%%%%%%
%
%
Substituting in Eq.~\eqref{Dirac2} the ansatz
%
%
%%%%%%%%%%%%
$$
\phi_r(u)= u^{\kappa_\text{i}/2} (1-u)^{\kappa_\text{e}/2} w_r(u)\, ,
$$
%%%%%%%%%%%%
%
%
we find the hypergeometric equation
%
%
%%%%%%%%%%%%
\begin{align}
u(1-u) w''_r + \left[ (\kappa_\text{i}+1)- (2\bar \kappa +2) u  \right] w'_r - 
\left[ 
\left( \bar \kappa + r s\tau \Delta_0  \right) 
\left( \bar \kappa - r s\tau \Delta_0 + 1 \right) 
\right] w_r =0\, .
\end{align}
%%%%%%%%%%%%
%
%
We need to select the solution which leads to normalizable states for
$y\rightarrow +\infty$, i.e., $u\rightarrow 0$. We find
%
%
%%%%%%%%%%%%
\begin{subequations}
\begin{align}
w_+(u) &=c_+\,  F[\bar \kappa +s\tau\Delta_0,
\bar \kappa -s\tau \Delta_0+1 ; 
\kappa_\text{i} +1; u]\, ,\\
w_-(u) &=c_-\, F[\bar \kappa - s\tau \Delta_0,
\bar \kappa + s\tau \Delta_0 +1 ; \kappa_\text{i} +1; u]\, ,
\end{align}
\end{subequations}
%%%%%%%%%%%%
%
%
where $F[a,b;c;z]$ is the ordinary hypergeometric function~\cite{Olver:2010}. 
Notice that  $F[a,b;c;z]=F[b,a;c;z]$.
The other solution to the hypergeometric equation does not lead to 
normalizable states and we omit it. Since we consider a semi-infinite system, 
in contrast to Ref.~\cite{Tchoumakov_2017}, we do not require normalizability 
for $y\rightarrow -\infty$, but we need to impose the appropriate boundary condition at $y=0$, 
as discussed in the main text. 
The relative factor between the two components is  fixed by the Dirac equation, and we find
%
%
%%%%%%%%%%%%
$$
\frac{c_-}{c_+} = \frac{\kappa_\text{i} +s\tau \Delta_\text{i}}{\tau k_x+E}= 
\frac{\tau k_x-E}{\kappa_\text{i}-s\tau \Delta_\text{i}} \, .
$$
Then, to summarize, up to an overall normalization factor, we have
%
%
%%%%%%%%%%%%
\begin{subequations}
\label{zigzagapp}
\begin{align}
\phi_+ &=  (\tau k_x+E) u^{\kappa_\text{i}/2} 
(1-u)^{\kappa_\text{e}/2} 
F[  \bar \kappa + s\tau\Delta_0 ,\bar \kappa - s\tau\Delta_0 + 1 ; \kappa_\text{i} +1; u]\, ,\\
\phi_- &= (\kappa_\text{i}+s\tau \Delta_\text{i}) u^{\kappa_\text{i}/2} 
(1-u)^{\kappa_\text{e}/2} 
F[\bar \kappa - s\tau\Delta_0 , \bar \kappa + s\tau\Delta_0+1; \kappa_\text{i} +1; u] \, .
\end{align}
\end{subequations}
%%%%%%%%%%%%
%
%
We observe that, since for $y\rightarrow +\infty$ we have $u \sim e^{-2(y-y_0)} \rightarrow 0$, 
and $F[a,b;c;u]\rightarrow 1$,
the decay of the wave functions in Eq.~\eqref{zigzagapp} is controlled by the parameter $\kappa_\text{i}$,
which can then be identified with the inverse decay length of the corresponding edge state.

%%%%%%%%
\subsection{Armchair case}

In the case of a semi-infinite system with an armchair edge along the $y$-axis, 
after factorization of a plane wave in the $y$-direction with wave vector $k_y$, 
the Dirac equation reads
\begin{align}
\left[-i \tau \sigma_x \frac{d}{dx} +\sigma_y k_y   + \sigma_z s\tau \Delta(x)\right] \psi(x) =E\psi(x)\, .
\label{Diracarm}
\end{align}
%%%%%%%%%%%%
%
%
Squaring Eq.~\eqref{Diracarm}, we obtain
%
%
%%%%%%%%%%%%
\begin{align}\label{Dirac3}
\left[ \frac{d^2}{dx^2} + (E^2-k^2_y-\bar \Delta^2- \Delta^2_0) + 
\frac{\Delta_0(\Delta_0 + s \sigma_y)}{\cosh^2(x-x_0)} -
2 \bar \Delta \Delta_0 \tanh(x-x_0) \right] \psi(x) =0\, .
\end{align}
%%%%%%%%%%%%
%
%
In this case, we express $\psi$ as
%
%
%%%%%%%%%%%%
$$
\psi =\left( \begin{array}{c}
v_A\\v_B
\end{array}\right) = \phi_+ |+\rangle + \phi_- |-\rangle, 
\qquad v_{A}=\frac{1}{\sqrt{2}} \left( \phi_+ + \phi_-\right)\, , \qquad 
v_B=\frac{i}{\sqrt{2}} \left( \phi_+ - \phi_-\right)\, ,
$$
%%%%%%%%%%%%
%
%
where $|\pm \rangle$ denote now the eigenvectors of $\sigma_y$, with respective eigenvalue $\pm1$.
Following the same steps as in the previous subsection, we find
%
%
%%%%%%%%%%%%
\begin{subequations}
\label{armchairapp}
\begin{align}
\phi_+ &=  d_+ u^{\kappa_\text{i}/2} 
(1-u)^{\kappa_\text{e}/2} 
F[  \bar \kappa - s\Delta_0 ,\bar \kappa + s\Delta_0 + 1 ; \kappa_\text{i} +1; u]\, ,\\
\phi_- &=d_- u^{\kappa_\text{i}/2} 
(1-u)^{\kappa_\text{e}/2} 
F[\bar \kappa + s\Delta_0 , \bar \kappa - s\Delta_0+1; \kappa_\text{i} +1; u] \, ,
\end{align}
\end{subequations}
%%%%%%%%%%%%
%
%
\end{widetext}
with the prefactors given by 
%
%
%%%%%%%%%%%%%
\begin{align}
\frac{d_-}{d_+}= \frac{E-k_y}{\tau (s\Delta_\text{i} +\kappa_\text{i})} = 
\frac{\tau (s\Delta_\text{i} -\kappa_\text{i})}{E + k_y} \, .
\label{prefactorarmapp}
\end{align}
%%%%%%%%%%%%
%
%
% 

We mention in passing that the solution for the armchair case can also be obtained by 
an appropriate $\pi/2$-rotation of the solution for the zigzag case.
Notice that  while the overall phase of the wave function is immaterial, 
the relative phase between $\phi_+$ and $\phi_-$ is important. 
The equation~\eqref{prefactorarmapp} implies that 
either $\phi_+$ or $\phi_-$ has opposite signs at the two valleys, 
while the other has the same sign. This observation turns out to be important 
when we impose the boundary condition, as discussed in Sec.~\ref{analyticsarmchair}.

%%%%%%%
%%%%%%%
\section{On the units conversion between tight-binding model and LWA results}
\label{app_parameters}
%%%%%%%%%%%%%%%%%%%%%%%%%%%%%%%
As already mentioned in the main text, the SOC parameters in the 
continuum Dirac equation and in the numerical tight-binding model 
are related as 
$$
\Delta = 3 \sqrt{3} \lambda\, .
$$ 
In the tight-binding model, we measure energy in units of the hopping amplitude $t$, 
length in units of the lattice constants $a_0$, 
and the Fermi velocity is given by $v_F = \frac{\sqrt{3}}{2} \frac{a_0 t}{ \hbar}$.
In the continuum, we measure energy in units of $\hbar v_F/\ell$, and wave vectors in 
units of $\ell^{-1}$. Then, the conversion formulas are
%
%
%
%%%%%%%%%%%%
\begin{align}
E_\mathrm{tb} &= \frac{E}{t} =  \frac{E_\mathrm{c}\frac{\hbar v_F}{\ell}}{t}= 
\frac{\sqrt{3}}{2} \frac{a_0}{\ell} E_\mathrm{c} = 
%\frac{\sqrt{3}}{24} E_\mathrm{c} =
0.072 E_\mathrm{c}\,, \nonumber \\
k_\mathrm{tb} & = ka_0 = k\ell \frac{a_0}{\ell} =% \frac{1}{12} k_\mathrm{c} = 
0.083 k_\mathrm{c}\, ,\nonumber
\end{align}
%%%%%%%%%%%%
%
%
where we have inserted the value $\ell=12a_0$ used throughout this paper. 
The linear relation $E= \hbar v_Fk$ in continuum units becomes $E_\mathrm{c}=k_\mathrm{c}$,
and in tight-binding units becomes
$$
E_\mathrm{tb} = \frac{ \sqrt{3}}{2} k_\mathrm{tb} = 0.87 k_\mathrm{tb}\,.
$$
%%%%%%%%%%%%
%
%
%%%%%%
\subsection{Zigzag case}
%%%%%%%%% 

With the zigzag boundary along $x$, we have
%
%
%%%%%%%%%%%%
$$
\Delta(y) =  \bar \Delta + \Delta_0\tanh \left(\frac{y-y_0}{\ell}\right) \, ,
$$
%%%%%%%%%%%%
%
%
where the continuum coordinate $y$ is given by
%
%
%%%%%%%%%%%%
$$
y=\sqrt{3}n a_0\, , \qquad n\in \mathbb{Z}\, ,
$$
%%%%%%%%%%%%
%
%
and $y_0=a_0/\sqrt{3}$. Since $x=na_0$ ($n\in \mathbb{Z}$), the 
one-dimensional Brillouin zone in the transport direction is $0<k_x< 2\pi/a_0$.

%%%%%%
\subsection{Armchair case}
%%%%%%%%%%%%

With the armchair boundary along $y$, we have
%
%
%%%%%%%%%%%%
$$
\Delta(x) =  \bar \Delta + \Delta_0\tanh \left(\frac{x-x_0}{\ell}\right)\, ,
$$
%%%%%%%%%%%%
%
%
with the continuum coordinate 
%
%
%%%%%%%%%%%%
$$
x=na_0, \qquad n\in \mathbb{Z}\, ,
$$
%%%%%%%%%%%%
%
%
and $x_0=a_0/2$. In this case, the continuum coordinate in the transport
direction is $y=\sqrt{3}n a_0$,
thus the corresponding Brillouin zone is $|k_y|<\frac{\pi}{\sqrt{3}a_0}$. 
In the plots showing the tight-binding band structure, however, the wave vectors are 
rescaled in such a way that the one-dimensional 
Brillouin zone appears to be $|k_\mathrm{tb}|<\pi$,
see Fig.~\ref{fig_BandsArmchair}.
Therefore, when comparing continuum and tight-binding results, it is important to 
take into account this additional $\sqrt{3}$ rescaling factor. 
In particular, the linear dispersion $E=\hbar v_Fk$
appears in the numerical results as 
%
%
%%%%%%%%%%%%
$$
E_\mathrm{tb} = \frac{1}{2} k_\mathrm{tb}\, .
$$
%%%%%%%%%%%%
%
%

%%%%%%%%%%%%%%%%%%%%%%% BAGWELL PHYSICS
\section{Conductance minima due to bound states}
\label{app_ConductanceDips}
%
%
%%%%%%%%%%%%
\begin{figure}[!h]
    \centering
    \includegraphics[width=\columnwidth]{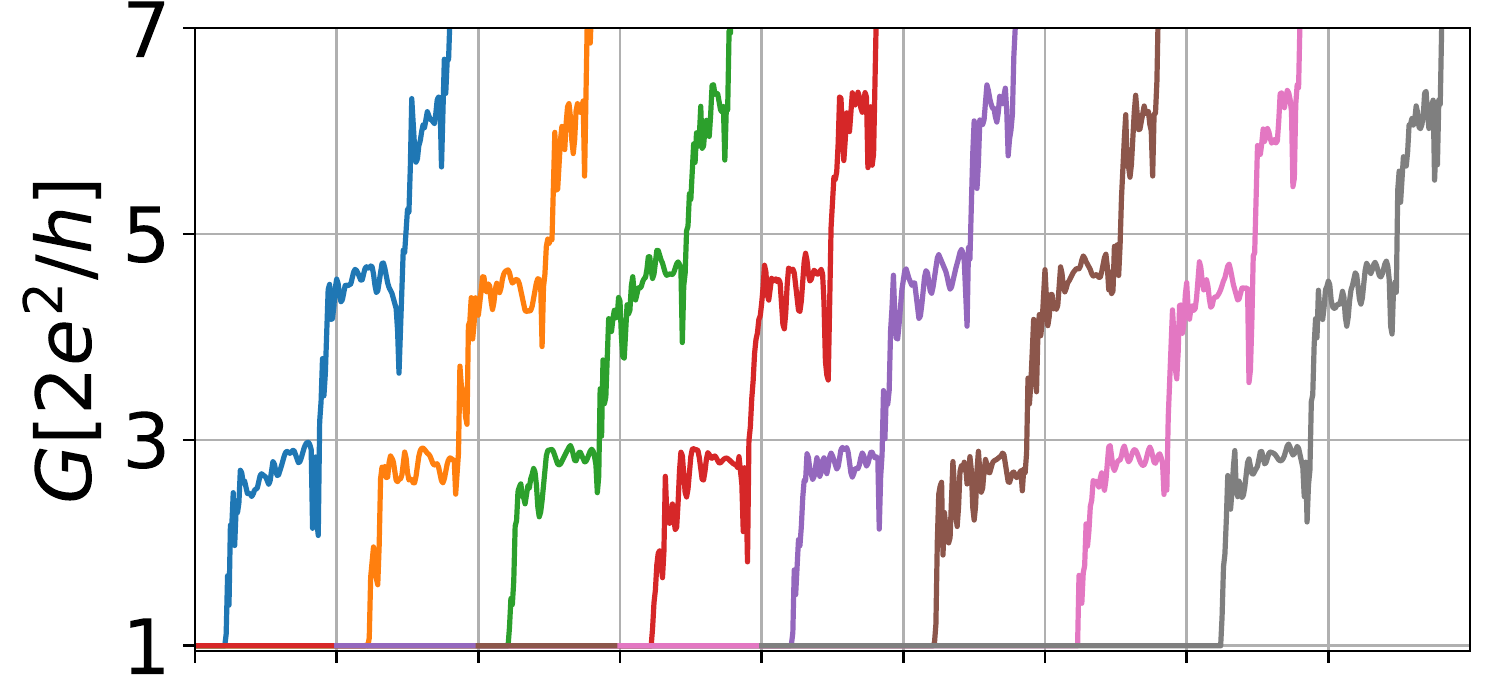}
    \caption{\label{fig_GSingleConfigs} Conductance curves for single disorder configurations in the armchair case for $U_0=0.1t$. 
    At the conductance plateaus,  random fluctuations eventually cancel each other out, 
    whereas the dip near the conductance step is in common to most disorder configurations, therefore it does not average out.}
\end{figure}
%%%%%%%%%%%%
%
%

In Fig.~\ref{fig_GSingleConfigs} we show conductance curves of single disorder configurations. It can be seen that 
the dip at the conductance step is in common to most realizations. This is because most disordered energy landscapes have 
room for quasi-bound states within certain local energy wells. These states, lying at energies just under the opening of the next VP 
mode, couple to the propagating mode, which therefore localises at that energy, thereby decreasing the conductance.

\begin{figure}[!t]
    \centering
    \includegraphics[width=\columnwidth]{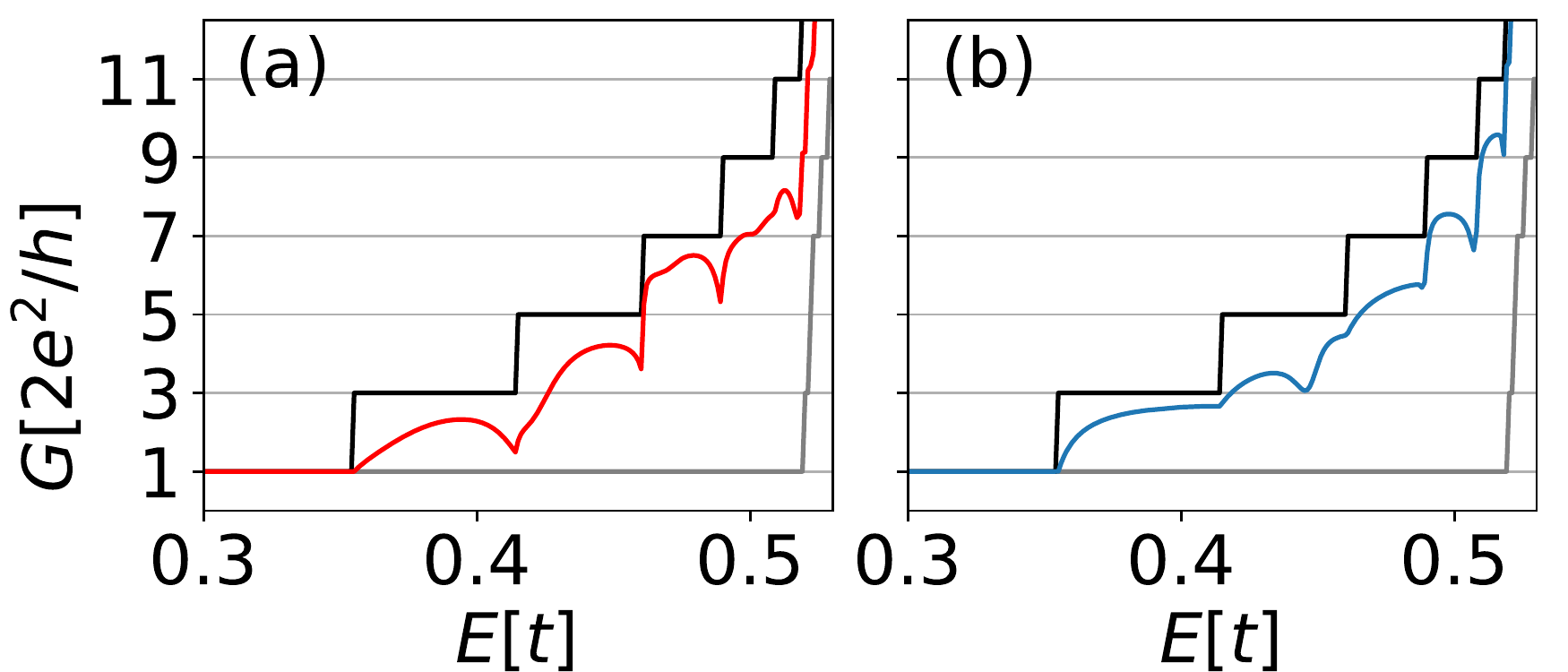}
    \caption{\label{fig_BagwellDip} Conductance for a clean system containing one Gaussian shaped impurity on each edge. In (a) $\sigma=3 a_0$ and in (b) $\sigma = a_0$, with $u_0=-0.4 t$ and positions $x_L =x_R=L_x/2$, and $y_L=8.848 a_0$ and $y_R=L_y-y_L$. In (a), due to the width of the impurity, all propagating VP modes couple to the evanescent mode at the impurity. In (b), because the impurity potential is much narrower, certain propagating modes do not interact with the evanescent mode at the impurity. }
\end{figure}

To have a qualitatively better understanding of what causes the dips that we systematically observe in the conductance curves for a disordered system, we have performed additional numerical simulations. Suspecting these minima are due to the coupling of the VP modes to quasi-bound states in the system, we run simulations in the more conventional setting of a clean nanoribbon with a single attractive impurity. In this case, we put a single impurity on each edge, shifted away from the edge by the same amount. This makes the impurity on each edge interact with the same VP mode. The impurity has a Gaussian shape 
%
%
%%%%%%%%%
\begin{align*}
U_0(\bm \zeta ) = u_0 \, e^{-\frac{(\bm \zeta - \bm \zeta_\text{L})^2+(\bm \zeta -\bm \zeta _\text{R})^2}{2 \sigma}},
\end{align*}
%%%%%%%%%
%
%
centered at $\bm \zeta _\text{L} = (x_\text{L},y_\text{L}) $ on one edge and $\bm \zeta _R = (x_\text{R},y_\text{R})$ on the other edge. The impurity is therefore fully characterised by its position, strength $u_0<0$, and width $\sigma$ (or variance $\sigma^2$). 
%The impurity is attractive, so $u_0 < 0$. 
As discussed in the main text, we indeed observe dips in the conductance near the steps. How many dips we see, at which steps, and their shape, depend on the properties of the impurities. For impurities that are very narrow, such as $\sigma = a_0$, we observe dips only at certain steps, and not at others, depending on where the impurity lies | c.f. Fig.~\ref{fig_BagwellDip}(b). For wider impurities, such as $\sigma=3a_0$, we observe dips at all steps, as well as other dips in the plateaus | c.f. Fig.~\ref{fig_BagwellDip}(a). Additional dips can result from having multiple evanescent modes at the impurity due to its finite width, or geometrical resonance effects. For the case of quasi-one dimensional quantum wires, it is also known that in the presence of Rashba SOC~\cite{Bercioux2015} the dips never go all the way down to the level of the previous conductance step, but are lifted proportional to the forth order in the Rashba SOC parameter~\cite{Alba_2024aa}. 
A systematic study of all these features is left to future investigations.
%
%
%%%%%%%%%%%%
%%%%%%%%%%%%
%
%

%%%%%%%%%%%%%%%%%%%%%%
% Bibliography
\bibliography{2dmaterialsQSH}

\end{document}